\documentclass[aps,floatfix,twocolumn,nofootinbib,preprintnumbers]{revtex4}
\usepackage[latin1]{inputenc}
\usepackage{amsmath}
\usepackage{amsfonts}
\usepackage{hyperref}
\usepackage{amssymb}
\usepackage{graphicx}
\usepackage{subfigure}
\usepackage{caption}
\usepackage{slashbox}
\usepackage{tabularx}
\usepackage{dcolumn}	
\usepackage{float}
\usepackage{bm}
\usepackage{latexsym}
\usepackage{epsfig}
\usepackage{graphicx,multirow}
\usepackage{color,soul}
\graphicspath{ {figures/} }

\newcommand{\cg}[6]{\mathcal{C}^{#1 #2}_{#3 #4 #5 #6}} 
\newcommand{\ncap}{\hat{n}}       

\newcommand{\be}{\begin{equation}}
\newcommand{\ee}{\end{equation}}
\newcommand{\ba}{\begin{eqnarray}}
\newcommand{\ea}{\end{eqnarray}}
\newcommand{\nn}{\nonumber \\}

\newcommand{\specialcell}[2][c]{%
  \begin{tabular}[#1]{@{}c@{}}#2\end{tabular}}

\begin{document}

\title{Orthogonal BipoSH measures : Scrutinizing sources of isotropy violation}

\author{Saurabh Kumar$^{1}$, Aditya Rotti$^{1}$, Moumita Aich$^{2}$, Nidhi Pant$^{1}$, Sanjit Mitra$^{1}$ and Tarun Souradeep$^{1}$}
\address{ $^1$IUCAA, Post Bag 4, Ganeshkhind, Pune-411007, India\\
$^2$Astrophysics and Cosmology Research Unit, School of Mathematics,
Statistics \& Computer Science, University of KwaZulu-Natal, Durban,
4041, South Africa\\
E-mail:  saurabhk@iucaa.ernet.in, aditya@iucaa.ernet.in, aich@ukzn.ac.za, nidhip@iucaa.ernet.in, sanjit@iucaa.ernet.in, tarun@iucaa.ernet.in}

\date{\today}

\begin{abstract}
The two point correlation function of the CMB temperature anisotropies is generally assumed to be statistically isotropic (SI). Deviations from this assumption could be traced to physical or observational artefacts and systematic effects. Measurement of non-vanishing power in the BipoSH spectra is a standard statistical technique to search for isotropy violations. Although this is a neat tool allowing a blind search for SI violations in the CMB sky, it is not easy to discern the cause of isotropy violation using this measure. In this article, we propose a novel technique of constructing orthogonal BipoSH estimators, which can be used to discern between models of isotropy violation. 


\end{abstract}
\maketitle
\section{Introduction.}
The Cosmic Microwave Background (CMB)  temperature anisotropy measurements are one of the cleanest probes of cosmology. The CMB full sky temperature anisotropy measurements have been used to test the assumption of the isotropy of the universe. The study of full sky maps from the WMAP 5 year data \citep{komatsu,spergel,nolta}, WMAP 7 year data \citep{wmap7-anomalies} and the recent PLANCK data \citep{planck23}, have led to some intriguing anomalies that may be interpreted to indicate deviations from statistical isotropy (SI). Deviations from SI can be caused by a number of physical, observational and systematic effects, such as, non-trivial cosmic topology \citep{TS}, Bianchi models \citep{jaffe, pontzen, ghosh}, primordial magnetic fields \citep{durrer, kahniashvili}, anisotropic primordial baryon-photon distribution \citep{MA-TS}, weak gravitational lensing \citep{AR-MA-TS}, asymmetric power spectrum \citep{pullen-kamionkowski}, modulation of the CMB sky \citep{planck23} and measurement of statistically isotropic (SI) CMB skies with non-circular 
beam \citep{hanson-beams-2010, das-beams-2014, joshi-das2012} being a small subset of the possibilities. 

The CMB temperature anisotropies are assumed to be Gaussian, which is in good agreement with current CMB observations. Hence the two point correlation function contains complete information about the underlying CMB temperature field. The generic two point correlation function can be completely encoded in the Bipolar Spherical Harmonic (BipoSH) basis. The coefficients of expansion in this set of basis function are termed as BipoSH spectra $A^{LM}_{l_1 l_2}$. While the BipoSH spectra with    $L=0$ encode information of the SI part of the correlation function, the rest of the BipoSH spectra detail the deviations from isotropy. Measuring non-vanishing BipoSH spectra $(A^{LM}_{l_1 l_2} ~;~ L \neq 0)$ forms the basic criteria behind searches for deviations from SI in CMB maps. 


In the search for deviations from SI, one has to use either one the following two strategies,
\begin{itemize}
\item Search for deviation from SI by measuring deviations from zero in BipoSH spectra. This method has the advantage of being model independent but non-optimal as it lacks sensitivity.
\item Search for deviation from SI by constructing an optimal estimator by resorting to a chosen model of SI violation. This method has the advantage of being optimal at the cost of being biased, as it requires working with a specific model of isotropy violation. 
\end{itemize}
The study discussed in this article presents a strategy to combine the advantages presented by the above two strategies, while minimising their drawbacks.

\section{Introduction to the BipoSH formalism} \label{sec1}
The general two point correlation function can be expressed in terms of the spherical harmonic coefficients of CMB temperature maps,
\ba
C(\mathbf{\ncap_1},\mathbf{\ncap_2})&=& \langle\Delta T(\mathbf{\ncap_1})\Delta T(\mathbf{\ncap_2})\rangle \\ 
	&=& \sum_{lml'm'} \langle a_{lm}a^*_{l'm'}\rangle Y_{lm}(\mathbf{\ncap_1}) Y_{l'm'}^{*}(\mathbf{\ncap_2})	\,.  \nonumber
\ea
In the SI case, this correlation function depends only on the angular separation between the two directions and not on the directions $\mathbf{\ncap_1}$ and $\mathbf{\ncap_2}$ explicitly. This property makes it possible to expand the correlation function in the Legendre polynomial ($P_l$) basis,
\ba
C(\mathbf{\ncap_1},\mathbf{\ncap_2})&=&C(\mathbf{\ncap_1} \cdot \mathbf{\ncap_2}) \\
     ~&=& \sum_l \frac{2l+1}{4\pi} C_{l} P_{l}(\mathbf{\ncap_1} \cdot \mathbf{\ncap_2}) \,, \nn
C_l~&=& \langle a_{lm}a^*_{l'm'}\rangle \delta_{l l'}\delta_{mm'} \,,
\label{unlensed-cl}
\ea
where $C_l$ is the standard angular power spectrum.

In the absence of SI, the correlation function does have an explicit dependence on the two directions $\mathbf{\ncap_1}$ and $\mathbf{\ncap_2} $. In this case, the BipoSH which form a complete orthonormal basis for functions defined on $S^2 \times S^2$, forms the ideal basis for studying the direction dependent two point correlation function \citep{AH-TS,AH-TS1}. The CMB two point correlation function is expressed in the BipoSH basis in the following manner,
\ba
C&
&\!\!\!\!\!\!\!\!\!\!\!\!(\mathbf{\ncap_1},\mathbf{\ncap_2})=\sum_{LMl_1l_2}
A^{LM}_{l_1 l_2} \left\{ Y_{l_1}
(\mathbf{\ncap_1})\otimes Y_{l_2}(\mathbf{\ncap_2})\right\}_{LM} \,,\\
	&=&\sum_{L M l_1 l_2} A^{LM}_{l_1 l_2} \sum_{m_1 m_2} \cg L M {l_1}
{m_1} {l_2} {m_2} Y_{l_1m_1} (\mathbf{\ncap_1})Y_{l_2 m_2}(\mathbf{\ncap_2})\,,
\nonumber
\ea
where $\cg L M {l_1} {m_1} {l_2} {m_2}$ are the Clebsch-Gordon coefficients, the indices of which satisfy the following triangularity relations $|l_1-l_2| \leq L \leq l_1+l_2$  and $m_1+m_2 = M$.

These BipoSH coefficients can be expressed in terms of the covariance matrix derived from observed CMB maps,
\ba \label{bips_cov}
 A^{LM}_{l_1  l_2} = \sum_{m_1 m_2} \langle a_{l_1 m_1}a_{l_2 m_2} \rangle \cg L
M {l_1} {m_1} {l_2} {m_2} \,.
\ea
In the case of SI the only non-vanishing BipoSH coefficients are $A^{00}_{l l}$ and they can be expressed in terms of the CMB angular power spectrum $C_l$,
\be \label{iBipoSH}
	A^{00}_{ll}=(-1)^l \Pi_l C_l, 	
\ee
where $\Pi_l=\sqrt{2l+1}$. 

\subsection*{A convenient notation  : $A^{LM}_{l_1l_2} \rightarrow A^{LM}_{l l + D}$}
We introduce the re-indexed BipoSH coefficients $A^{LM}_{l l + D}$ which are more convenient to interpret. One can now think of the indices  $M \in \{-L,L\}$ \& $D \in \{-L,L\}$ 
as independent parameters with $l$ representing the inverse angular scale of the CMB map. 
\section{Constructing orthogonal BipoSH measures} \label{construct-orth-biposh-measure}

Many isotropy violation mechanisms \citep{pullen-kamionkowski, AR-MA-TS, joshi-das2012} can be shown to give rise to BipoSH spectra of the following form,
\be \label{general-iso-vio}
A^{LM}_{l l + D} = (-1)^l \Pi_l C_{l} \delta_{L0}\delta_{M0}\delta_{D0} + p_{XY} G^{L}_{l l + D} \,,
\ee
where $C_l$ is the angular power spectrum, $G^{L}_{l l + D}$ denotes the template shape function arising from any model of isotropy violation which in may be expressed as a function of $C_l$ and $p_{XY}$ denotes the parameter detailing the isotropy violation, where the parameter indices can map either to $ \{X,Y\} \equiv \{L,M\}$ or $ \{X,Y\} \equiv \{l,l\} $.

One can now estimate the parameter $p_{LM}$ by inverting Eq.~\ref{general-iso-vio} to yield the following estimator,
\be
\hat{p}_{XY}= \frac{A^{LM}_{l l + D}}{G^{L}_{l l + D}} \,.
\ee
Each value of the indices $l~ \&~ D$ provide an independent estimate of $p_{XY}$. 

It is common practice to make an optimal minimum variance sum to extract the best estimate of $p_{XY}$. In this article, we draw attention to an alternate construct in which we take an optimal minimum variance difference of the independent estimates of $p_{XY}$. The basic idea behind this construction is to check for consistency between parameter values as estimated from independent BipoSH coefficients. If the model for isotropy violation is appropriate, then it predicts the correct template shape function $G^{L}_{l l + D}$ encoded in the independent BipoSH coefficients, hence yielding the same parameter value irrespective of the BipoSH coefficient used to evaluate it. On the contrary, if the model for isotropy violation is invalid, then it predicts a wrong shape function encoded in the different BipoSH coefficients, leading to a discrepant values of the parameter as evaluated from different BipoSH coefficients. \textit{Therefore assessing the statistical significance of the differences between parameter values as extracted from independent BipoSH coefficients can be used as a self consistency test to rule 
out models of isotropy violation.}

We propose the following estimator as a self consistency test for isotropy violation models,
\ba \label{orthogonal-biposh-estimator}
\hat{d}_{XY} &=& \hat{p}_{XY} -\hat{p}^\prime_{XY} \,, \nonumber \\
&= &\sum\limits_{l_{\mathrm{min}}}^{l_{\mathrm{max}}} w^L_{l} \left[\frac{A^{LM}_{l l + D}}{G^L_{l l+D}}-\frac{A^{LM}_{l l+D^\prime}}{G^L_{l l+D^\prime}}\right] \,,
\ea
where $w_l$ are the weights used to minimise the variance on the estimator. It can be shown that this change in sign makes no difference to the statistics of the estimator, as the independent BipoSH modes are uncorrelated. Generally for any model of isotropy violation which can be cast in the form of Eq.~\ref{general-iso-vio} it can be shown that the weights that minimise the variance are given by the following expression,
\be
w^L_{l} = \frac{\left[\frac{C_{l} C_{l + D} (1+\delta_{D0})}{\left(G^{L}_{l l+ D}\right)^2}+\frac{C_{l}C_{l+D^\prime}(1+\delta_{D^\prime0})}{\left(G^{L}_{l l+D^\prime}\right)^2}\right]^{-1}}{\sum\limits_{l}^{l_{\mathrm{bin}}} \left[\frac{C_{l} C_{l + D} (1+\delta_{D0})}{\left(G^{L}_{l l+ D}\right)^2}+\frac{C_{l}C_{l+D^\prime}(1+\delta_{D^\prime0})}{\left(G^{L}_{l l+D^\prime}\right)^2}\right]^{-1}} \,.
\ee

Note that this is similar to the minimum variance estimator used in the case of lensing reconstruction \cite{okamoto-hu}, except that the plus sign is replaced with a minus sign in Eq.~\ref{orthogonal-biposh-estimator}. It can be also be shown that the sensitivity of the orthogonal BipoSH estimator is given by the following expression, 
\be
N_l^L = \frac{1}{\sum\limits_{l}^{l_{\mathrm{bin}}} \left(\frac{C_{l} C_{l + D} (1+\delta_{D0})}{\left(G^{L}_{l l+ D}\right)^2}+\frac{C_{l}C_{l+D^\prime}(1+\delta_{D^\prime0})}{\left(G^{L}_{l l+D^\prime}\right)^2}\right)^{-1}} \,,
\ee
where $N^L_l$ in the  variance of the orthogonal BipoSH estimator for an isotropic CMB sky whose statistical properties are described by the angular power spectrum $C_l$.
 
\section{Testing models of statistical isotropy violation} \label{sec2}

In the following section we discuss some models of isotropy violation and the template shape functions arising from them. Following which, we demonstrate the ability of the orthogonal BipoSH estimators to discern between models of isotropy violation by evaluating them on ideal full sky, non-SI simulated CMB maps. Finally we present the results of an identical study carried out on WMAP observed maps.

\subsection{Sources of isotropy violation} 
\label{ideal-case-study}
\textbf{Weak lensing : }Since any given realisation of large scale structure surrounding an observer is anisotropic, it imprints a signature of isotropy violation in the lensed CMB sky. Weak lensing of the CMB photons results in the CMB temperature map getting remapped as described by the following equation,
\ba \label{remap}
\tilde{T}(\ncap)&=& T(\ncap+\vec{\nabla} \psi(\ncap)) \,,
\ea
where $\tilde{T}$ denotes the lensed CMB temperature anisotropies, $T$ denotes the unlensed CMB temperature anisotropies and $\psi$ is the projected lensing potential. 

Making no assumptions about isotropy of the lensed CMB sky and evaluating the two point correlation in the BipoSH basis results in the following expression,
\ba \label{lensbips}
\tilde{A}^{LM}_{l l+D} &= & (-1)^l \Pi_l C_{l} \delta_{L0}\delta_{M0}\delta_{D0} + \psi_{LM} G^L_{l l+D} \,,
\ea
where the lensing shape function $G^L_{l l+D}$ is expressed in terms of the CMB angular power spectrum as follows,
\ba \label{lensing-shape}
G^L_{l l+D} &=& \frac{C_{l}F(l+D,L,l)+C_{l+D}F(l,L,l+D)}{\sqrt{4\pi}} \nn &\times& \frac{\Pi_{l} \Pi_{l+D}}{ \Pi_{L}}~\cg L 0 {l} 0 {(l+D)} 0 \,,
\ea
where,
\ba
F(l_1,L,l_2)&=&\frac{\left[ l_2(l_2+1)+L(L+1)-l_1(l_1+1)\right]}{2} \,.
\nonumber
\ea
\\
%


\textbf{Modulation : } Modulation of the CMB sky is another model which violates SI. This is being used as a phenomenological model to explain the anomalous, large angular scale dipolar asymmetry seen in the observed CMB maps \cite{wmap7-anomalies,planck23}. Modulated CMB temperature anisotropy map is mathematically expressed as follows,
\be \label{modulate-cmb}
\tilde{T}(\ncap)=T(\ncap) \left[ 1+ P(\ncap) \right] \,,
\ee
where $P$ denotes the modulating field.

Expressing the two point correlation function for the modulated temperature field in BipoSH coefficients results in an equation having the following form,
\be
\tilde{A}^{LM}_{l l+D} = (-1)^l \Pi_l  C_{l} \delta_{L0}\delta_{M0}\delta_{D0} + P_{LM} G^L_{l l+D} \,,
\ee

where $P_{LM}$ are the spherical harmonic coefficients of the modulating field, while the modulation shape function $G^L_{l l+D}$ is expressed in terms of the CMB angular power spectrum as follows,

\be
G^L_{l l+D} = \frac{C_{l}+C_{l+D}}{\sqrt{4\pi}} \frac{\Pi_{l} \Pi_{l+D}}{ \Pi_{L}}~\cg L 0 {l} 0 {(l+D)} 0 \,.
\ee
\\
%


\textbf{Anisotropic power spectrum : } Another model of isotropy violation is where the primordial fluctuations are described by a direction dependent power spectrum \cite{AP-MK, YZM-GE-AC, LA-SC-MW},
\begin{eqnarray} \label{dd-pps}
P(\vec{k})&=&\mathcal{P}(k) \left[1+ g(\hat{k})\right] \,, \\ &=& \mathcal{P}(k) \left[1+ \sum_{LM}g_{LM} Y_{LM}(\hat{k})\right] \nonumber
\end{eqnarray}
where $\mathcal{P}(k)$ describes the standard isotropic power spectrum while $g(\hat{k})$ encodes the directional dependence of the statistical properties of the primordial density fluctuations.

It can be shown that a direction dependent PPS results in the following set of BipoSH coefficients,
\begin{equation} \label{biposh-from-dd-pps}
A^{LM}_{l l+D} = (-1)^l \Pi_l C_{l} \delta_{L0}\delta_{M0}\delta_{D0} + g_{LM} G^L_{l l+D} \,,
\end{equation}
where $g_{LM}$ are the spherical harmonic coefficients of the direction dependent part of the primordial power spectrum and the shape function introduced due to the anisotropic power spectrum is given by the following expression,
\begin{eqnarray} \label{shapefn-from-dd-pps}
G^L_{l l+D} &=&  \frac{i^D}{\sqrt{4 \pi}} \frac{\Pi_{l} \Pi_{l+D}}{\Pi_L} \mathcal{C}^{L0}_{l 0 l+D 0} \\ &\times& \int \frac{2 k^2 dk}{\pi} \Delta_{l}(k)\Delta_{l+D}(k) P(k) \,. \nonumber
\end{eqnarray}
Note that the shape function $G^{L}_{l l+D}$ can be expressed in terms of the angular power spectrum $C_l$ in the case where $D=0$, however the rest of the coefficients with $D \neq 0$ require explicit evaluations of the integral in Eq.~\ref{shapefn-from-dd-pps}. \\


\textbf{Non-circular beam : } It is known that measurement of CMB temperature anisotropies by non-circular instrument beams results in the observed sky being non-SI \citep{Mitra2004, AH-TS}. This has been realised as one of the most prominent systematics which needs to be accounted for, while searching for deviations from SI in the CMB sky.

Most generally the measured CMB temperature anisotropies are related to the true CMB sky through the following expression,

\begin{equation}
\label{cmb-beam-convolve}
\tilde{T}(\hat n)=\int d{\hat n}^{'} B(\hat n,{\hat n}^{'}) T({\hat n}^{'}) \,,
\end{equation}

where $\tilde{T}(\hat n)$ represents the measured CMB sky, $B(\hat n,{\hat n}^{'})$ denotes the beam sensitivity function and $T({\hat n}^{'})$ represents the true CMB sky. The beam function is characterised as being circular if it satisfies : $B({\hat n}, {\hat n}^{'}) = B(\cos^{-1}({\hat n} \cdot{\hat n}^{'}))$, while any deviations from this condition render it non-circular. 

It can be shown that, mildly non-circular beams result in the generation of the BipoSH spectra \cite{joshi-beams-2012}, which can be expressed in the following form,
\begin{equation}\label{ncbeambiposh}
\tilde A^{L M}_{l l+D} =  (-1)^l \Pi_l C_{l} B_{l}^2 \delta_{L0}\delta_{M0}\delta_{D0} + b_{l2} G^{L}_{l l+D} \,,
\end{equation}
where $B_{l}$ beam transfer function which characterises the circular part of the beam, while $b_{l2}$ are the beam spherical harmonics ($b_{lm} \textrm{ with } m=2$) pointed along the north pole $(\hat{z})$, which characterise the non-circularity of the beam and $G^{L}_{l l+D}$ is the  induced shape function which can be expressed as follows,
\begin{eqnarray} 
G^{L}_{l l+D} &=& \frac{2\pi\Pi_L}{\left(\Pi_{l l+D}\right)C^{L0}_{l 0 l+D 0}} \\ &\times& \Big[C_{l+D}B_{l+D} \xi^{L0}_{l l+D} + C_{l} B_{l}\xi^{L0}_{l+D,l}\Big] \nonumber
\label{ncbeam-shape}
\end{eqnarray}
where,
\begin{eqnarray}
\xi^{L0}_{l_{1}l_{2}} &=&\frac{\Pi_{l_{1}}}{\sqrt{(4\pi)}}\sum_{m_{2}}(-1)^{m_2}C^{L0}_{l_1 -m_2 l_2 m_2} \nonumber \\  &\times&  \int^{\pi}_{0} d^{l_2}_{m_2 2}(\theta)d^{l_1}_{m_2 0}(\theta)\sin\theta d\theta\
\end{eqnarray}
In arriving at Eq.~\ref{ncbeambiposh}, we have made a number of assumptions. We assume the beam non-circularity to be small which allows us to retain terms only up to first order in the beam non-circularity parameter ($b_{lm}$) and truncate modes above $| m |=2$ for all the multipoles $l$. Further to be able to arrive at an semi-analytically evaluable expression for the non-circular beam induced shape function we assume a constant scanning strategy. Specifically in the case of WMAP, it has been shown that \cite{joshi-beams-2012} the beam non-circularity and the detailed scanning strategy is well approximated by an effective beam and a constant scan strategy (i.e. any chosen axes along the beam maintains a constant angle with a reference coordinate on the sphere). 

An interesting point to note is that all the source models of isotropy violation under study, modulation, weak lensing by large scale structure, anisotropic primordial power spectrum and non-circular beam effects, only induce even parity (i.e. $D+L$ is even)  BipoSH coefficients.  Explicitly this behaviour is due to the presence of $\cg L 0 {l} 0 {(l+D)} 0$ which vanishes when $D+L$ is odd. Though not considered here, the construction of orthogonal BipoSH estimator is equally valid even to models which generate odd parity BipoSH spectra.

\subsection{Testing the orthogonal estimators on simulated SI violating maps}
\label{sec2a}

\noindent Given the shape function induced due to any particular source of SI violation, it is possible to arrive at an orthogonal BipoSH estimator, following the  construction discussed in Section~\ref{construct-orth-biposh-measure}. To test and demonstrate the effectiveness of these estimators we evaluate them on a set of simulated SI violating CMB maps. For this example study we consider the following two cases,
\begin{itemize}
\item A modulated CMB maps, constructed as in Eq.~\ref{modulate-cmb}, where we have used a modulation field of the form $P(\ncap) = 0.1 Y_{20}(\ncap)$.
\item A non-circular beam convolved CMB map. These maps are constructed following the same procedure as described in \cite{SD-SM-AR-NJ-TS}. We use the side A and side B of the W1 beam of WMAP for the convolution process and use a realistic WMAP scanning strategy.
 \end{itemize}
We evaluate the BipoSH estimator and the orthogonal BipoSH estimators for weak lensing, modulation, anisotropic power spectrum \& non-circular beams on these simulated maps. 

Before discussing the results of our analysis on these simulated maps, we reiterate that the orthogonal BipoSH estimates yield difference between model parameter estimates as evaluated from independent BipoSH coefficients. \textit{Higher the significance of the orthogonal BipoSH estimates, the more likely it is for the source model under consideration to be invalid.} 

\begin{figure} [htbp]
\centerline{\psfig{figure=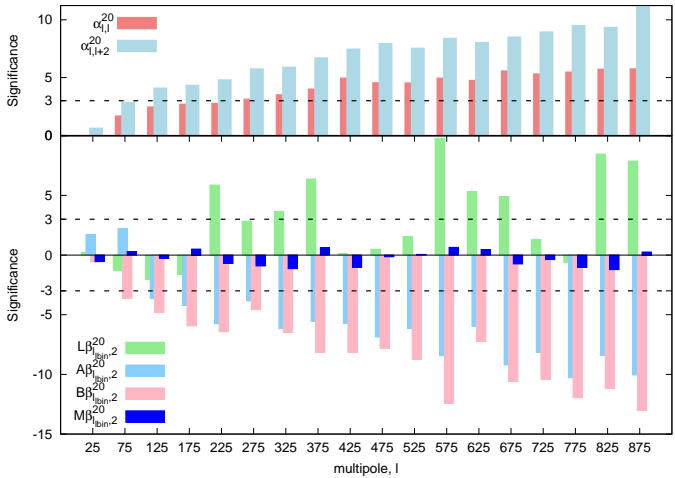, width = 6.5 cm, angle=-90}}
\captionsetup{singlelinecheck=off,justification=raggedright}
\caption{ \textbf{Top panel}: The red and the blue bars denote the significance of detection of the BipoSH spectra evaluated from simulated modulated maps constructed using a modulation field given by $P(\hat{n})=0.1 Y_{20}(\hat{n})$. \\
\textbf{Bottom panel}: The bars denote the significance of the difference in model parameters as derived from independent BipoSH modes. The green, light blue and pink  bars representing the sources, lensing (L), anisotropic power spectrum (A) and non-circular beam (B) respectively, show highly discrepant model parameters. Note that, the dark blue bars which represent modulation (M) model are seen to be consistent with zero within $3 \sigma$. }
\label{robust1}
\end{figure}

\begin{figure} [htbp]
\centerline{\psfig{figure=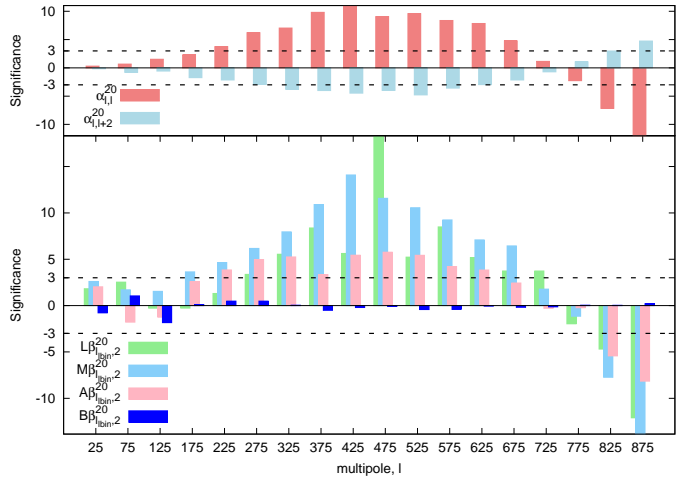, width = 6.5 cm, angle=-90}}
\captionsetup{singlelinecheck=off,justification=raggedright}
\caption{ \textbf{Top panel}: The red and the blue bars denote the significance of detection of the BipoSH spectra evaluated from a non-circular beam convolved map. \\
\textbf{Bottom panel}: The bars denote the significance of the difference in model parameters as derived from independent BipoSH modes. The green, pink and light blue bars representing the sources, lensing (L), anisotropic power spectrum (A) and modulation (M) respectively, show highly discrepant model parameters. Note that, the dark blue bars which represent non-circular beam (M) model are seen to be consistent with zero within $3\sigma$.}
\label{robust2}
\end{figure}

On evaluating these orthogonal estimators on modulated CMB maps, it is seen that all other source models, i.e. weak lensing, anisotropic power spectrum and non-circular beam, show high significance detections in the orthogonal BipoSH estimates, implying that these models yield statistically significant discrepancy in the same model parameter as estimated from independent BipoSH modes. However in the case when the source model assumed is modulation, the orthogonal BipoSH estimates are found to have extremely low significance as compared to other models. This means that the parameter for the assumed source model (modulation field harmonics in this case), as evaluated from independent BipoSH modes, are consistent with each other within error bars, as seen in Fig.~\ref{robust1}. Similarly when these estimators are evaluated on CMB maps convolved with non-circular beam, the discrepancy between model parameters is found to be least significant only when the source model assumed is that of non-circular beam, as 
seen in Fig.~\ref{robust2}, while all the other source models under consideration yield high discrepant values for the same model parameter.

The results from this exercise are summarised in Table.~\ref{robust1}, where we quote the cumulative significance of the model parameter discrepancy as evaluated from independent BipoSH modes. The most favoured model i.e the source model showing the least discrepant model parameters is highlighted.
\begin{table}[!h] 
\setlength{\tabcolsep}{2pt}
\renewcommand{\arraystretch}{1.0}
\begin{center}
\begin{tabular}{|c|c|c|c|c|}
\hline
\backslashbox{Simulation}{Source}
& \specialcell{Weak\\lensing} & \specialcell{Anisotropic\\PPS} & \specialcell{NC\\beam} & Modulation\\
\hline
Modulation & $64.1$ &  $112.1$ & $141.9$ & $\textcolor{red}{11.0}$\\
\hline
NC beam & $92.6$ &  $66.3$ & $\textcolor{red}{7.1}$ & $122.7$\\
\hline
\end{tabular}
\end{center}
\vskip -0.3 cm
\captionsetup{singlelinecheck=off,justification=raggedright}
\captionof{table}{This table represent the cumulative significance of the difference between the model parameter estimates as derived from independent BipoSH modes. While the rows indicate the nature of the non-SI map used for the study, the columns indicate the source model assumed for the analysis. The source model which yields the least significant orthogonal BipoSH estimate is highlighted.}
\label{robust-table}
\end{table}

Through these example case studies we have demonstrated that the orthogonal BipoSH estimators can be used to quantitatively assess the most favoured model of isotropy violation given the data. 

\subsection*{Revisiting the WMAP quadrupolar anomaly}  \label{sec3}
\begin{figure} [htbp]
\centerline{\psfig{figure=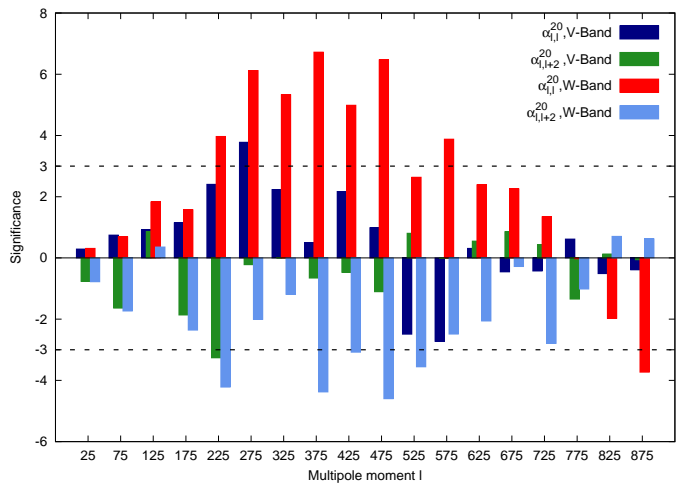, width = 6.5 cm, angle=-90}}
\captionsetup{singlelinecheck=off,justification=raggedright}
\caption{This plot depicts the significance of BipoSH spectra from WMAP 9 foreground reduced temperature maps (\href{www.lambda.gsfc.nasa.gov}{LAMBDA} site). We evaluate the BipoSH spectra in the ecliptic coordinate system and use KQ75 temperature mask to negotiate with foregrounds. Note that there are highly significant detections in $A_{l l}^{20}$ and $A_{l l+2}^{20}$ modes in both V-Band and W-Band.}
\label{WMAP9yrBips}
\end{figure}
In this section we address the following question : \emph{Given the source models of isotropy violation : weak lensing, modulation, non-circular beam and anisotropic power spectrum, can the orthogonal BipoSH estimators be used to assess, which of the models provides the most viable explanation for the WMAP quadrupolar anomaly \citep{wmap7-anomalies}  ?}
\begin{figure}[htbp] 
  \centering
  \subfigure{\includegraphics[height=8.5cm,width=6cm, angle=-90]{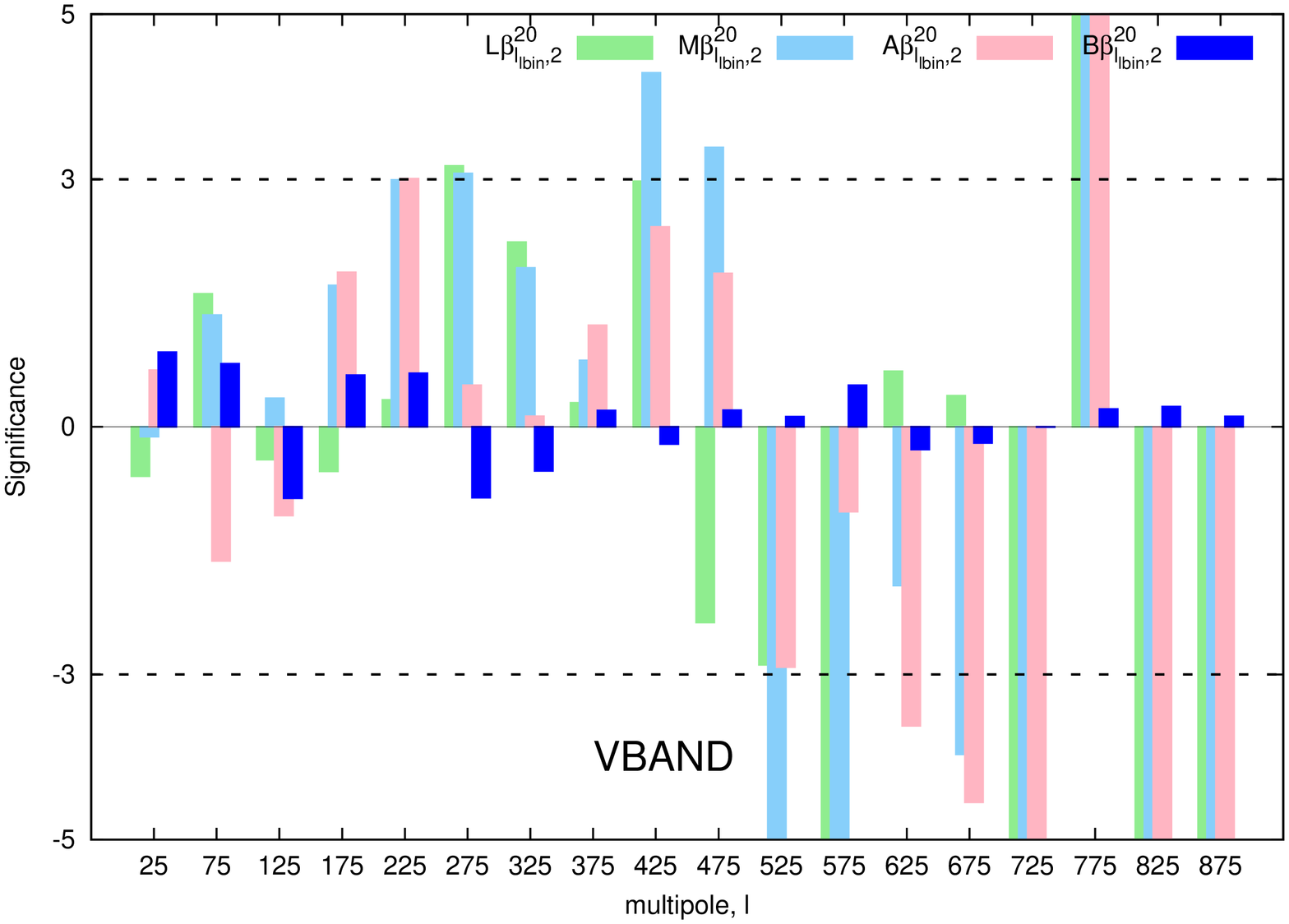}}
  \subfigure{\includegraphics[height=8.5cm,width=6cm, angle=-90]{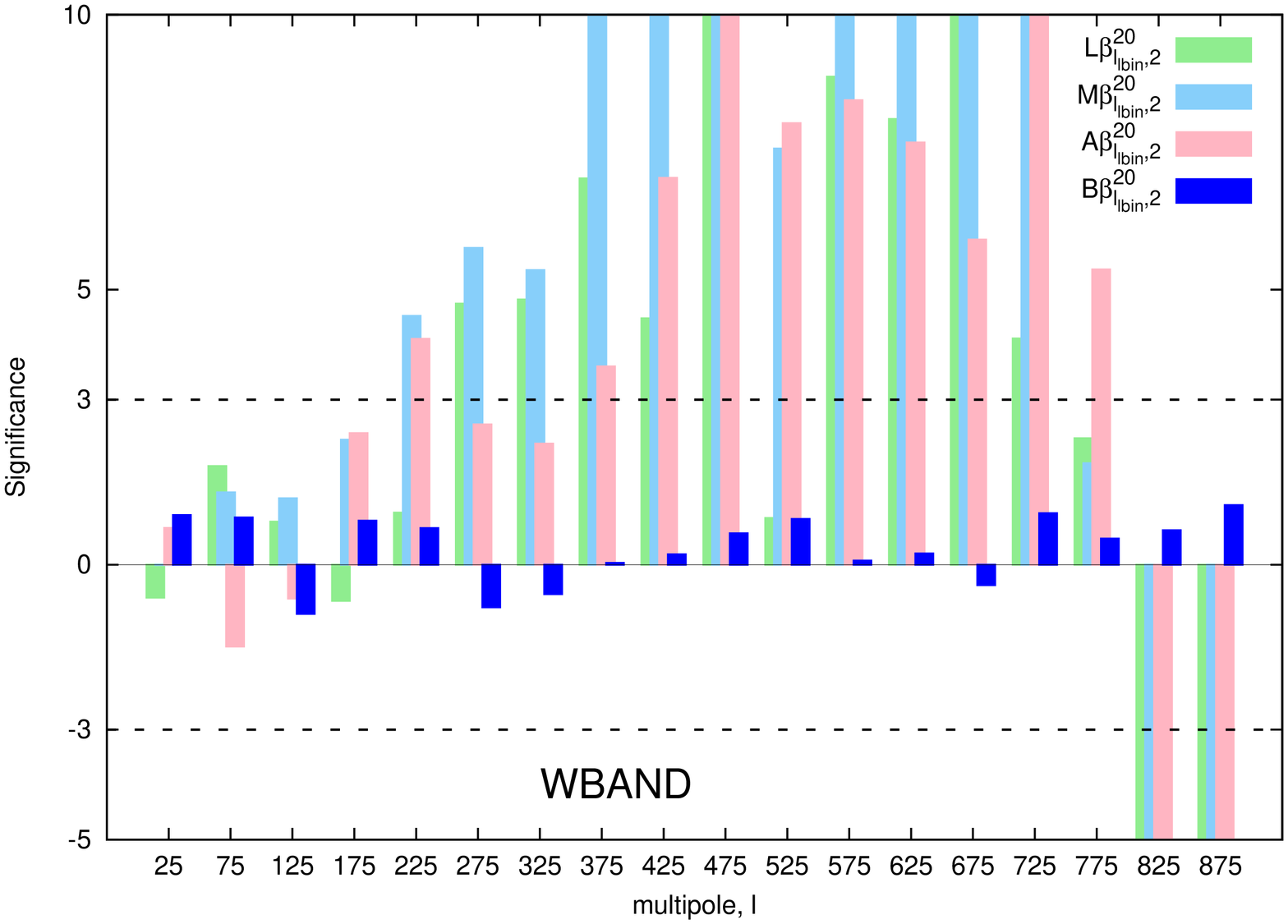}}
  \captionsetup{singlelinecheck=off,justification=raggedright}
  \caption{ The green bars denote the significance of the parameter difference defined via Eq.~\ref{orthogonal-biposh-estimator} for the lensing independent estimator, while the pale-blue and pink bars denote the same for the modulation and anisotropic power spectrum independent estimators respectively. To retain spectral information, the orthogonal BipoSH measures were evaluated by computing the minimizing variance sum over multipoles of bin-width $\Delta l = 50$.}
\label{compare-model-si-sources}
\end{figure}
Unlike in the case of the example case studies presented in Section~\ref{sec2a}, where we dealt with a CMB sky with no noise, no foreground contaminations and no sky cuts, real data is invariably plagued by these systematics. The biases introduced due to these known systematics need to be carefully accounted. While evaluating the orthogonal BiposH estimator, we follow the analysis procedure described in the work \citep{AR-MA-TS}, which attempted at explaining the WMAP quadrupolar anomaly as arising from weak lensing of the CMB photons.

We first evaluate the BipoSH spectra from WMAP maps and evaluate the significance of the detections as depicted in Fig.~\ref{WMAP9yrBips}. Recall that detection of non-vanishing BipoSH spectra indicate a violation of SI. 
\begin{table}[!h]
\setlength{\tabcolsep}{2pt}
\renewcommand{\arraystretch}{1.0}
\begin{center}
\begin{tabular}{|c|c|c|c|c|}
\hline
\backslashbox{Band}{Source}
& \specialcell{Weak\\lensing} & \specialcell{Anisotropic\\PPS} & \specialcell{NC\\beam} & Modulation\\
\hline
V & $94.2$ &  $71.4$ & $\textcolor{red}{7.5}$ & $121.9$\\
\hline
W & $141.3$ &  $117.7$ & $\textcolor{red}{10.8}$ & $188.4$\\
\hline
\end{tabular}
\end{center}
\vskip -0.3 cm
\captionsetup{singlelinecheck=off,justification=raggedright}
\captionof{table}{This table represent the cumulative significance of the difference between the model parameter estimates as derived from independent BipoSH modes. The most preferred model is one for which this difference is least significant. The difference between the non-circular beam parameters is found to be least significant as compared to other source models of SI violation, indicating that the effects of non-circular beam is the most likely source for the signal seen in WMAP measurements of the CMB sky. The lower cumulative significance seen for V-band is indicative of the lower significance of BipoSH spectra detections themselves.}
\label{table_VW}
\end{table}
Finally we evaluate the orthogonal BipoSH estimators in order to converge on the most viable explanation for these highly significant detections seen in the WMAP maps.  The results of the orthogonal BipoSH estimation analysis on V-band and W-band are presented in Fig.~\ref{compare-model-si-sources}. 
It is found that the orthogonal BipoSH estimates for the source models namely, lensing, modulation and anisotropic power spectrum show strong deviations ($> 3\sigma$) from nullity, in many CMB multipole bins, clearly indicating that these model provide a poor explanation for the BipoSH spectra detections seen in WMAP maps. On the contrary for the case of the model of non-circular beam induced SI violation, it is seen that the orthogonal BipoSH estimates are consistent with nullity within error bars, for all the CMB multipole bins, unambiguously pointing to the most favoured model. We have also summarised this plot in the Table~\ref{table_VW}, where we quote the cumulative significance of deviation from nullity for all the orthogonal BipoSH estimators. It is found that the cumulative significance of deviation from nullity is expectedly minimal in the case of non-circular beam induced anisotropy as compared to any other model of isotropy violation under consideration.

These findings are in complete synchrony with existing studies which addressed these detections in WMAP data and have thoroughly established that these detections were indeed due to unaccounted non-circular beam effects \citep{joshi-beams-2012, planck23, das-beams-2014, hanson-beams-2010}.

\section{Discussion} \label{sec4}

In this article we have proposed the novel orthogonal BipoSH estimators which can be used to discern between models of isotropy violation. These estimators can be used as a self consistency test for any model trying to explain non-SI signatures seen in the data. 
We have given a general prescription for constructing such orthogonal BipoSH estimators for any model of isotropy violation which can be cast in the form of Eq.~\ref{general-iso-vio}. This happens to be the case in many popular models of SI violation like, modulation of the CMB sky \citep{planck23}, anisotropic primordial power spectrum \citep{pullen-kamionkowski}, weak lensing and asymmetric beams induced SI violations \citep{joshi-das2012}.

We have constructed the orthogonal BipoSH estimators for the above mentioned models and carried out a systematic study of these estimators on ideal (full sky \& no noise) non-SI simulated maps. Through this systematic study we have established that this method can be used to quantitatively discern between models of isotropy violation and converge to the most preferred model given the data. 

Finally we have carried out a similar analysis on WMAP data. Our study reveals that the most preferred cause for the WMAP quadrupolar anomaly is that of non-circular beam effects. This only reconfirms in an independent fashion this well known result. This exercise serves to demonstrate the usability of these techniques on real data.

We acknowledge the use of HEALPix package \citep{healpix}. AR acknowledges the Council of Scientific and Industrial Research (CSIR), India for financial support (Grant award no.
20-6/2008(II)E.U.-IV). MA, SK and TS acknowledges support from the Swarnajayanti
fellowship, DST, India.

\bibliographystyle{plain}
\bibliography{ref.bib}

\end{document}